\documentclass[10pt,twocolumn,preprintnumbers,superscriptaddress,nofootinbib,aps,prl]{revtex4-2}

\usepackage{graphicx}
\usepackage{amsmath}
\usepackage{srcltx}
\usepackage[utf8]{inputenc}
\usepackage[colorlinks]{hyperref}
\usepackage{times}
\usepackage{amssymb}
\newcommand{\vev}[1]{\langle {#1} \rangle}
\newcommand{\lsim}{\lesssim}
\newcommand{\gsim}{\gtrsim}

\newcommand{\eq}[1]{Eq.~(\ref{#1})}

\newcommand{\ord}[1]{\mathcal{O}{(#1)}}
\newcommand{\beq}{\begin{equation}}
\newcommand{\eeq}{\end{equation}}
\newcommand{\bea}{\begin{eqnarray}}
\newcommand{\eea}{\end{eqnarray}}
\newcommand{\eps}{\varepsilon}

\newcommand{\mP}{M_{\rm P}}

\newcommand{\appropto}{\mathrel{\vcenter{
  \offinterlineskip\halign{\hfil$##$\cr
    \propto\cr\noalign{\kern2pt}\sim\cr\noalign{\kern-2pt}}}}}

\begin{document}

\pagestyle{plain}

\title{\boldmath First Order Electroweak Phase Transition from Weakly Coupled sub-GeV Physics
\\ and\\ 
Possible Connection to Fermion Flavor}

\author{Hooman Davoudiasl}
\email{hooman@bnl.gov}


\affiliation{High Energy Theory Group, Physics Department \\ Brookhaven National Laboratory,
Upton, NY 11973, USA}


\begin{abstract}

We propose that the dynamics of a scalar $\phi$ of mass $\ord{10}$~MeV that is weakly coupled to the Higgs can lead to a first order electroweak phase transition, fulfilling a key requirement for  baryogenesis.  Stability of the model near the weak scale requires a suppressed - possibly vanishing - top Yukawa coupling to the Higgs before the transition which rises to the Standard Model value afterwards.  This can be accomplished through the dynamics of $\phi$ via a dimension-5 operator.  We conjecture that the entire Standard Model flavor structure could turn on, {\it mutatis mutandis}, after the electroweak phase transition, via dimension-5 interactions of $\phi$ suppressed by scales ranging from $\ord{10^3}$~TeV to near Planck mass.  Due to its suppressed couplings, $\phi$  is long-lived and can lead to missing energy signals in rare kaon decays, which can be probed by the KOTO experiment.   

\end{abstract}
\maketitle


\section{Introduction}

It is widely accepted that the Standard Model (SM) does not contain sufficient ingredients to generate the observed baryon asymmetry of the universe (BAU) \cite{Tanabashi:2018oca}.  This motivates introduction of new interactions that can contribute to a successful baryogenesis mechanism.  Among the outstanding problems of particle physics and cosmology, the origin of the BAU could be a good target for laboratory experiments, since it concerns the {\it visible} world.  

One of the necessary conditions for viable baryogenesis is departure from equilibrium, in order to 
avoid erasure of the generated BAU \cite{Sakharov:1967dj}.  This condition could have been provided through a first order electroweak phase transition in the SM \cite{Cohen:1990py}.  However, the measured mass of the Higgs boson, 
$m_H \approx 125$~GeV \cite{Tanabashi:2018oca}, rules out this possibility.  
The basic reason is that at this mass, given the well-established vacuum expectation of the Higgs $\vev{H} = v/\sqrt{2}\approx 174$~GeV, its deduced 
self-coupling $\lambda_H \approx 0.13$ is too large to accommodate a first order phase transition (FOPT).  

To see the reason for the above situation, note that a FOPT is typically assumed to 
be realized if at the transition critical temperature $T_c$ we have
\beq
\frac{v(T_c)}{T_c} \gsim 1\, .
\label{FOPT-cond}
\eeq  
In the SM, the above quantity is governed by the thermal contributions of the $\{W^\pm, Z\}$ gauge bosons and is given by \cite{Quiros:1999jp,Carena:2004ha}  
\beq
\frac{v(T_c)}{T_c} = \frac{2 m_W^3 + m_Z^3}{3 \pi \lambda_H \, v^3}\approx 0.1\,,
\label{SM-param}
\eeq
where $m_W \approx 80.4$~GeV and $m_Z \approx 91.2$~GeV \cite{Tanabashi:2018oca} are the masses of the $W^\pm$ and $Z$, respectively.  
Hence, the SM electroweak transition 
seems to be far from being first order and thus fails to provide the necessary non-equilibrium condition for successful baryogenesis. 

The above circumstance 
motivates extensions of the SM that could lead to a FOPT.  Generally speaking, such 
extensions require new physics close to the weak scale that has non-negligible interactions with the SM.  For example, additional bosons with $\ord{1}$ coupling strength to the Higgs can enhance the value of the transition parameter in \eq{SM-param} \cite{Nelson:1991ab,Espinosa:1993yi,Carena:1996wj,Kanemura:2004ch,Profumo:2007wc,Noble:2007kk,No:2013wsa,Katz:2014bha,Basler:2016obg,Kurup:2017dzf,Chen:2017qcz,Ramsey-Musolf:2019lsf,Basler:2019iuu}.  In that case, their masses could not be much larger than the Higgs mass, so that their thermal population is not Boltzmann-suppressed at $T_c\sim m_H$.  However, see Refs.~\cite{Jeong:2018ucz,Kozaczuk:2019pet} for mechanisms that effect a FOPT through scalars of intermediate mass $\gsim 10$~GeV and weak coupling with the Higgs.  A model that achieves an electroweak FOPT from a light axion-like-particle is presented in Ref.~\cite{Jeong:2018jqe}.      Ref.~\cite{Carena:2004ha} examines a model that employs TeV scale fermions.  See Ref.~\cite{Ghosh:2020ipy} for recent work using a dark gauge sector leading to a FOPT.

Another class of solutions for achieving a FOPT entails addition of higher dimension operators $(H^\dagger H)^n$, with integer $n>2$.  Such interactions would allow deviations from the SM value of $\lambda_H$ in order to satisfy the condition (\ref{FOPT-cond}) and obtain a first order transition \cite{Grojean:2004xa,Delaunay:2007wb}.  In this class of models, it is generally required that the higher dimension operators are suppressed by scales $\Lambda \sim 1$~TeV, which implies the presence of new physics not far above $m_H$.  For a realization of such effective theories using weak scale fermions see Ref.~\cite{Davoudiasl:2012tu}.  

We see that, in general, one is led to introduce new physics near the weak scale that couples to the Higgs with un-suppressed strength.  This situation can cause tension with a growing body of data that typically does not seem to favor the presence of the required states. 

In this work, we propose an alternative approach to generating an electroweak FOPT, where a light new scalar $\phi$, with suppressed couplings to the Higgs boson and other SM particles, appears at energies well below the weak scale.  The main role of the scalar $\phi$ is to allow for a small Higgs self-coupling before electroweak symmetry breaking, so that the ensuing transition is first order.  After the electroweak symmetry is broken, we arrange for the Higgs self coupling to attain its low temperature value, corresponding to the value deduced from experiment.  This is achieved by the vacuum expectation value (vev) of 
$\phi$ that becomes non-zero after the FOPT and drives the vev of the Higgs to $v\approx 246$~GeV,  observed at low energies. 

The above picture can generally lead to an unstable scalar potential just above the weak scale.  This is mostly due to the quantum effects of the top quark Yukawa coupling, $y_t\approx 1$ in the SM, that drive the Higgs quartic to negative values with increasing energy scale.  We hence additionally propose that the top Yukawa coupling can be small or vanishing before the transition, ensuring a quantum mechanically stable potential at energies well above the weak scale.  The non-zero $\phi$ vev 
attained after the transition is then postulated to set $y_t$ to its SM value, through a dimension-5 operator.  Motivated by symmetry considerations, we are led to conjecture that all SM flavor may be established in this fashion, through dimension-5 operators suppressed by mass scales ranging from $\sim 10^3$~TeV, for the top, to near the Planck mass $M_P\sim 10^{19}$~GeV, for neutrinos.          

The end result of our mechanism is that the Higgs potential and couplings after the FOPT are to a very good approximation those of the SM.  The suppressed interactions of $\phi$ imply that other new physical states are well above the weak scale or else have small couplings to the SM.  The only other field that we explicitly introduce is a scalar with a mass $\sim 10$~TeV after the transition and a small coupling to $\phi$, which is therefore largely inaccessible.  This typically leads to the only observable new effects being from $\phi$, which in our setup is a weakly coupled sub-GeV state.  See also Ref.~\cite{Baldes:2016gaf} for a model that contains a light scalar and relates flavor dynamics to an electroweak FOPT.     

Our proposal  would then generically be consistent with the lack of significant deviations at or near the weak scale, which is currently probed at the LHC and various precision experiments.  However, low energy probes of rare phenomena, such as certain kaon decay modes, could in principle provide experimental tests of our scenario.  Here, we note that if the FOPT generated in our proposal is to be part of a larger framework for baryogenesis, there would likely be other new states that couple to the SM and could gives rise to additional signals.  Such signals will depend on the specific features of baryogenesis models; 
we will come back to this point later.

We will next introduce an explicit model that realizes our FOPT scenario.

\section{Scalar Potential}

Consider the following potential  
\bea\nonumber
V(\phi, H, \eta) &=& \frac{m_{0\phi}^2}{2} \phi^2 - (\mu_0^2 + 2  \mu\, \phi) H^\dagger H \\ 
&+& 
(\lambda_0 + 2 \frac{\phi^2}{M^2})(H^\dagger H)^2 + \frac{\kappa}{4} \phi^2 \eta^2 \,,
\label{V}
\eea
where $m_{0\phi}^2>0$ is the initial mass of $\phi$, positive constants $\mu_0^2$ and $\lambda_0$ are the initial values of the Higgs mass parameter and self-interactions for $\vev{\phi} = 0$, respectively, and $0<\mu \ll m_H$.  The scale $M\gg m_H$ descends from some ultraviolet (UV) dynamics that we do not specify here.  We have included an additional scalar $\eta$ 
coupled to $\phi$ with positive  
strength $\kappa \ll 1$ whose role is to allow a thermal ``slow roll" for $\phi$, as will be described below.   The above potential is consistent with a softly broken 
${\mathbb Z}_2$ symmetry acting on $\phi$.

There can be other possible terms in the potential $V(\phi, H, \eta)$ that we may invoke later, but the above interactions suffice to elucidate key aspects of our mechanism, for now.  We will assume that other interactions not invoked in this work are suppressed compared to those that we explicitly write down here and below.  Also, we will only examine if the chosen parameters can yield a consistent phenomenology, regardless of whether they are considered tuned or not.  
In particular, a more complete model leading to our effective theory may be required to avoid potentially large quantum corrections to $\phi$ mass.  While we do not offer any concrete examples here, we point out that $\phi$ only needs to emerge in our theory below scales of $\Lambda\sim \ord{100~\text{GeV}}$ to affect the FOPT.  In that case, loop corrections to its mass could in principle be cutoff at $\Lambda$.  The largest contribution to $m_\phi\sim$~few MeV would be from interactions with $\eta$ in our effective description.  Assuming $\kappa\sim 10^{-4}$, as we will later, one can show the required tuning is $\ord{10^{-3}}$, for $\Lambda\sim 100$~GeV.

In what follows, we will assume that a FOPT can be achieved for $v(T_c)/T_c \sim 1$, in agreement with condition (\ref{FOPT-cond}).  Over the range of parameters that we will consider, we find that the values of the SM $SU(2)_L$ and $U(1)_Y$ hypercharge couplings, respectively $g$ and $g'$, would not run significantly and stay nearly constant.
As $m_W, m_Z \propto v$, \eq{SM-param} then implies that we will need $\lambda_0\lsim 0.1\, \lambda_H$ at the onset of the transition, to satisfy relation (\ref{FOPT-cond}). 

To find the requisite values of various parameters, we obtain the vacuum solutions for the scalars,  corresponding to $\partial_h V = \partial_\phi V=\partial_\eta V=0$, where $h$ is the background value of the Higgs boson, with $\vev{h}=v=246$~GeV.  We will set $\vev{\eta}=0$.  Let $\bar \phi$ denote the background value of $\phi$ at $T=0$.  We find
\beq
\bar\phi = \frac{\mu \,v^2\,M^2 }{m^2_{0\phi}M^2 + v^4}\quad\quad (T=0).
\label{phibar}
\eeq
Using \eq{V}, the SM Higgs mass parameter $\mu_H \approx 89$~GeV and self-coupling $\lambda_H \approx 0.13$ -- which we will assume as the effective low energy values -- are given by
\beq
\mu_H^2 = \mu_0^2 + 2 \mu\, \bar \phi \; \;\; \text{and}
\; \; \; \lambda_H = \lambda_0 +2 \frac{{\bar \phi}^2}{M^2}\,.
\label{SMvals}
\eeq 
Let us parameterize $\lambda_0 = \eps \lambda_H$, with $\eps\lsim 0.1$ a small quantity.  In what follows, we will choose $\eps$, $\mu$, and $M$ as our input parameters that will determine the required values of 
$\mu_0$ and $m_{0\phi}$, setting $h=v$ at $T=0$.  We demand $m_{0\phi}^2>0$ and $\mu_0^2 >0$, so that the potential is stable and that electroweak symmetry can be spontaneously broken, respectively.  Then, using Eqs.(\ref{phibar}) and (\ref{SMvals}), we obtain
\beq
\frac{\bar \phi \, v^2}{M^2} < \mu < \frac{\mu_H^2}{2\bar \phi}\,.
\label{mu-range}
\eeq

\section{Mixing with the Higgs}

The mixing between the Higgs and $\phi$ is governed by the angle
\beq
\theta \approx \frac{2v^2}{m_H^2}\left(\frac{2\bar \phi\, v}{M^2} - \frac{\mu}{v}\right)\,.
\label{theta}
\eeq 
We will work in the regime $|\theta| \ll 1$ and hence the mixing effect on the Higgs can be ignored, which yields $v= \mu_H/\sqrt{\lambda_H}$.  After diagonalizing the mass matrix, to a good approximation, we will have $m_H \approx \sqrt{2} \,\mu_H$, as in the SM.  In that limit, the lightest scalar eigenstate, which we will continue to denote by $\phi$, will end up having a mass given by 
\beq
m_\phi^2 \approx m_{0\phi}^2+ \frac{v^4}{M^2} - \theta^2\, m_H^2\,,
\label{mphi2}
\eeq
that we will require to be positive, to ensure a valid solution for $\phi$.  

Let us take $\eps=0.1$ and $M= 4.0\times 10^3$~TeV as benchmark values, for concreteness.  Using Eqs.~(\ref{theta}) and (\ref{mphi2}), one can then obtain a 
range of values for $m_\phi$ and $\theta$ as a function of $\mu$.  By numerical inspection of \eq{mphi2}, we find that $\phi$ mass varies by a factor of $\sim 2$ over the range (\ref{mu-range}) and we have 
\beq
5~\text{MeV} \lsim m_\phi \lsim 10~\text{MeV}\,, 
\label{mphi-range}
\eeq
given our benchmark parameters.

For the region of parameters typical of our benchmark choices, the lifetime $\tau_\phi$ of $\phi$ corresponds to a macroscopic decay length  
\beq
c\,\tau_\phi \approx \frac{8\pi}{\theta^2 y_e^2 m_\phi} \approx 
8.4 \times 10^3~\text{km}
\left(\frac{10^{-8}}{\theta^2}\right)\! \! \left(\frac{7~\text{MeV}}{m_\phi}\right)\,,
\label{tauphi}
\eeq
where $c$ is the speed of light and $y_e \approx 2.9 \times 10^{-6}$ is the SM Yukawa coupling of the electron.  The above formula is valid for $\phi$ masses below the muon pair threshold.  Later, we will introduce an additional $\phi$ coupling to the electron, which however will not drive the $\phi$ lifetime away from the above order of magnitude, set by $\theta y_e$.   

\section{Evolution with Temperature}

Here, we will provide an estimate for the expected critical temperature $T_c$ of the FOPT.  One could obtain this quantity from the following expression (see, for example, Refs.~\cite{Quiros:1999jp,Carena:2004ha}) 
\beq
T_c \approx \frac{T_0}{\sqrt{1 - E^2/(\lambda_0\, D)}}\,
\label{Tc}
\eeq
where 
\bea
&&T_0^2 = \frac{m_{0H}^2 - 8 B v_0^2}{4 D}\; ; \; B = \frac{3(2 m_W^4 + m_Z^4 - 4 m_t^4)}{64 \pi^2 v^4}\nonumber \\
&&D = \frac{2 m_W^2 + m_Z^2 + 2 m_t^2}{8 v^2}\; ; \; 
E = \frac{2 m_W^3 + m_Z^3}{6\pi v^3}\,.
\label{BDET02}
\eea

In the above, $T_c$ corresponds to the temperature when there are two degenerate minima, which allows a transition to begin.  Also, $m_{0H}=\sqrt{2}\, \mu_0$, and $v_0^2 = \mu_0^2/\lambda_0$, corresponding to the initial Higgs mass and vev, respectively, at the onset of the transition ($\vev{\phi}=0$).  The value of the field at the new minimum, which will eventually become the new vacuum, is given by $h(T_c) \approx 2 E T_c/\lambda_0$.  At temperature $T_0$, the barrier separating the two minima disappears and the origin at $h=0$ becomes a local maximum.  Hence, the transition is complete between $T_c$ and $T_0$.  The value of the Higgs field at this minimum is given by 
$h(T_0) \approx 3 E T_0/\lambda_0$. 

For the benchmark values of this work, and with $\mu$ in the range (\ref{mu-range}), we find that 
$0 \leq T_c \leq 95$~GeV, with $T_c=0$ at the upper limit of the range $\mu = \mu_H^2/(2 \bar \phi)\approx 4.1$~MeV.  We do not consider this endpoint to yield a sensible cosmology and hence will only consider 
\beq
3.7~\text{MeV} \leq\mu\leq 4.0~\text{MeV}\,, 
\label{mu-vals}
\eeq
as a possible range, corresponding to $38~\text{GeV} \leq T_c \leq 90$~GeV and $37~\text{GeV} \leq T_0 \leq 87$~GeV.   A narrower range can also be considered, if desirable.  Note that in the above, $m_t$ is the top mass at the onset of electroweak symmetry breaking, which we set $m_t=0$.  Later, we will discuss the reason for this choice, which is motivated by vacuum stability of the Higgs potential above the weak scale.  A mechanism for achieving the SM value after the FOPT will be provided below.  Thus, here we use $m_t = 0$ to obtain $T_c$ from the above expressions.

\begin{figure}[t]
\centering
\includegraphics[width=\columnwidth]{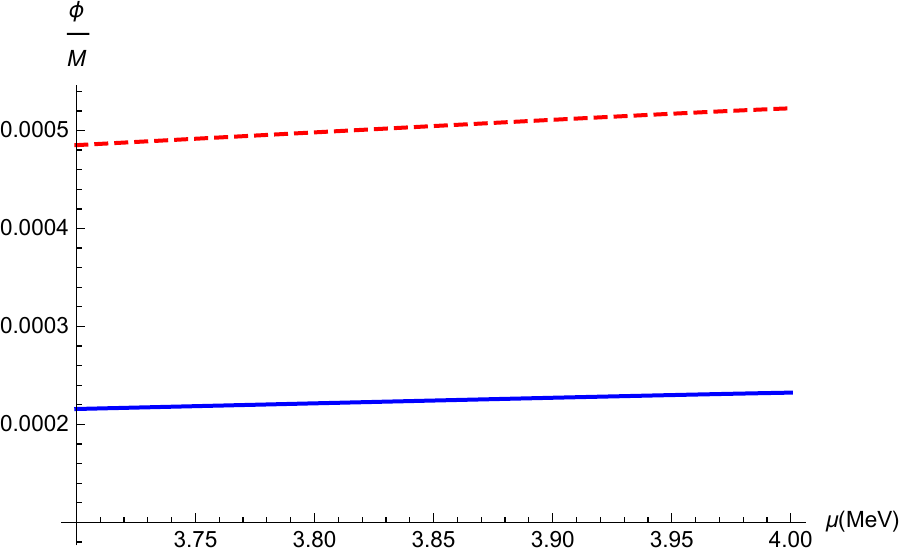}
\caption{Values of $\phi/M$ at $T_c$ (solid) and $T_0$ (dashed), assuming the benchmark parameters, from \eq{phiT}.  Here, $\kappa=10^{-4}$ has been assumed, for $\phi$-$\eta$ coupling in $V(\phi, H, \eta)$.  The smallness of $\phi/M$ implies that the Higgs parameters in \eq{SMvals} remain close to their initial values during the transition.}
\label{fig:phioverM}
\end{figure}

In order to make sure that the jump in the order parameter of the FOPT remains effective during the transition, the evolution of $\phi$ away from zero cannot be too prompt; this has been an implicit assumption in the preceding discussion.  In particular, if $\phi$ becomes large once $h$ tunnels to the new minimum, the Higgs quartic coupling could approach $\ord{1}$ values and the condition in \eq{FOPT-cond} may no longer hold.  
To see why this could be a problem, note that the Hubble scale around the time of the phase transition is roughly $T^2/\mP \sim 10^{-12}$~MeV, whereas the initial mass of $\phi$ is $\ord{\text{MeV}}$.  Hence, in the absence of some plasma effect, the evolution of $\phi$ could in principle be much faster than the relevant time scales for the completion of the transition.
To avoid this possibility, we will employ thermal effects from the interactions of $\phi$ with $\eta$, introduced earlier in \eq{V}.  The scalar $\eta$ generates a temperature dependence $\propto \kappa$ for $\phi$ mass \cite{Quiros:1999jp}   
\beq
m_\phi^2(T) = m_{0\phi}^2 + \frac{h^4(T)}{M^2} + \frac{\kappa}{24} T^2\,,
\label{mphiT}
\eeq
where $h(T)$ is the Higgs background value at temperature $T$.  

By inspecting the expression 
in \eq{phibar} for $\bar \phi$, we see that a sufficiently large thermal mass for $\phi$ will keep it pinned near $\phi =0$.  Once $T$ is low enough, $\phi$ can reach its late 
time values.  We have plotted values of 
\beq
\frac{\phi(T)}{M} = \frac{\mu \,h^2(T)}{m_\phi^2(T) M}\,,
\label{phiT}
\eeq
at $T_c$ (solid) and $T_0$ (dashed), in Fig.\ref{fig:phioverM}.  We have taken $\kappa=10^{-4}$ as a reference value.  As one can see, over the range (\ref{mu-vals}) of $\mu$ considered in this work, $\phi/M$ remains 
sufficiently small that one could assume, as implied by \eq{SMvals}, the Higgs parameters remain near their initial values $(\mu_0,\lambda_0)$ during the transition, to a good approximation.  Numerical inspection shows that $\phi/M$ remains small compared to unity down to $T\sim 10$~GeV.  
For these temperatures, we have checked that $\kappa\lsim \text{few}\times 10^{-5}$ yields $\phi/M\gsim 0.1$ where the Higgs quartic starts to become large compared to $\lambda_0 \sim 0.01$ and hence we do not consider $\kappa\ll 10^{-4}$, to be conservative.  

The delayed evolution of $\phi$ could possibly lead to a period of inflation which could result in  unwanted effects, such as dilution of any BAU generated during or before the FOPT.  To examine this question, we note that any such inflation may be caused if the potential energy stored in $\phi$ - that we estimate to be at most $\delta V\sim m_{0\phi}^2\, {\bar \phi}^2 \sim 10^6$~GeV$^4$ for our choice of reference parameters - dominates the cosmic energy budget.  
The radiation energy density is given 
by $\rho_R \sim g_* T^4$, where $g_*\sim 100$ is the number of relativistic degrees of freedom.  For $T\sim 10$~GeV, below which - as mentioned above - $\phi$ starts to evolve significantly towards the minimum of the potential, we find $\rho_R\sim 10^6$~GeV$^4$.  Thus, we see that such an inflationary period is not a typical expectation for the regime of parameters considered here.  

An implicit assumption in the above discussion is that $\phi$ and $\eta$ are in thermal equilibrium during the phase transition.  To see why this is the case, note that the $\phi$-Higgs coupling leads to $H H^\dagger \to \phi$ at a rate $\Gamma \sim \mu^2/T$, with $\mu \sim$~MeV.  The Hubble rate is roughly given by ${\cal H} \sim T^2/M_P$.  Demanding $\Gamma\gsim {\cal H}$, in order to populate a thermal bath of $\phi$ states, we obtain the condition $T\lsim (\mu^2 M_P)^{1/3}$.  For $\mu \sim$~MeV, we then find $T\lsim 10^{4}$~GeV.  Since $\eta$ is initially massless, its production via $\phi\phi\to\eta\eta$ has a rate $\sim \kappa^2 T$.  Hence, it can be in thermal equilibrium for $\kappa =10^{-4}$, used as a reference value above, as long as $T\lsim 10^{11}$~GeV.  We thus conclude that these considerations do not pose any restrictions on the FOPT - which only begins at $T_c\lsim 100$~GeV - and we can have $\phi$ and $\eta$ in thermal equilibrium with the SM.

We note that once the final values of the fields have been achieved, $\eta$ obtains a mass $m_\eta \sim \kappa^{1/2} \bar\phi \sim 10$~TeV, through its coupling to $\phi$.  Since we are assuming that 
$\eta$ has a thermal population in the early Universe, we need to make sure that its number density 
is sufficiently depleted, so that it would not lead to excessive contributions to dark matter.  One could in principle entertain the possibility that $\eta$ is dark matter, which would require 
arranging for suitable interactions that would yield an acceptable relic density for it.  We do not pursue 
that option here, though it would be an interesting complement to our model.  However, in lieu of such model building, one may simply postulate a dimension-5 operator, say $\eta F_{\mu\nu}F^{\mu\nu}/M_\eta$, with $F_{\mu\nu}$ the photon field strength tensor and $M_\eta$ some UV scale, to deplete the $\eta$ number density.  This operator would lead to $\eta \to \gamma \gamma$ sufficiently fast - that is before Big Bang Nucleosynthesis (BBN) at $T\sim$~MeV - as long as 
$M_\eta \lsim 10^{18}$~GeV, and would remove $\eta$ from the cosmic energy budget.        

\section{UV Stability of the Potential}

So far, we have shown that the proposed model can in principle yield a FOPT.  However, the central feature of our setup, namely a small initial quartic coupling $\lambda_0$ for the Higgs, can lead to an unstable potential, due to running which would drive $\lambda_0$ negative at large energy scales.  Denoting the top Yukawa coupling by $y_t$, the running of the initial Higgs quartic coupling is given 
by (see, for example, Ref.~\cite{Batell:2012zw}; note that this reference defines the Higgs quartic with an extra factor of 1/2 compared to our convention)
\bea\nonumber
16 \pi^2 \frac{d \lambda_0}{d t} &=& 24 \lambda_0^2 + 
\lambda_0 (12 y_t^2 - 9 g^2 - 3 g'^2) - 6 y_t^4\\ 
 &+& \frac{9}{8} g^4 +\frac{3}{8} g'^4 + \frac{3}{4} g^2 g'^2\,,
\label{running}
\eea
where $t \equiv \ln Q/Q_0$ and $Q$ is the renormalization scale; $Q_0$ is a reference scale.  We will set $Q_0=100$~GeV as the typical scale of the electroweak symmetry breaking.  As was discussed before, within the regime of parameters in this work, the critical  temperature $T_c \lsim 100$~GeV and hence this choice of $Q_0$ is reasonable.   

\begin{figure}[t]
\centering
\includegraphics[width=\columnwidth]{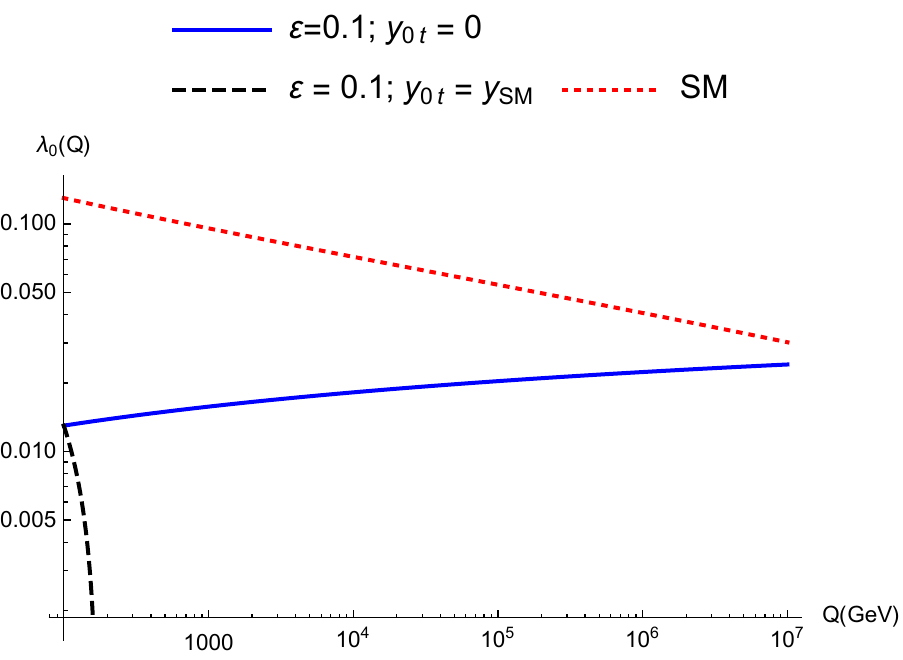}
\caption{Running of $\lambda_0$, for $\lambda_0(Q_0) = 0.1 \lambda_H$, corresponding to a top Yukawa coupling that is initially vanishing (solid) or at the SM value (dashed); for the latter choice the Higgs self-coupling goes negative at an energy scale $Q\gsim 170$~GeV.  The dotted curve corresponds to the running in the SM, with $\lambda_0(Q_0)=\lambda_H=0.13$.  The reference scale is chosen to be $Q_0=100$~GeV.}
\label{fig:running}
\end{figure}

Let us set $y_t(Q_0)=\sqrt{2}\,160 \,\text{GeV}/v\approx 0.92$ and the gauge couplings to their SM values near the weak scale.  For our benchmark choice, $\lambda_0(Q_0) \approx 0.013$, using \eq{running} we find that this parameter goes negative at energies above $\sim 170$~GeV, not far from the electroweak symmetry breaking temperature, as shown in Fig.~\ref{fig:running} by the dashed line.    This is largely due to the quantum effects of the top quark that couples to the Higgs with $\ord{1}$ strength and drives $\lambda_0$ to negative values.  Hence, as it is, the validity of the model above such energy scales requires the introduction new bosonic degrees of freedom that would stabilize the Higgs potential.  Such a solution, however, would lessen the motivation for our mechanism, since new weak scale bosons that couple to the Higgs with $\ord{1}$ stregngth could also provide the necessary ingredients for a FOPT, as has been studied extensively before.  As we will discuss below, the dynamics of $\phi$ itself offers a way to stabilize the Higgs potential without introducing additional weak scale bosons with significant coupling to the SM. 

To see how $\phi$ can help with the stability of the potential, first note that it is the large top Yukawa coupling that is the origin of the problem, as mentioned before.  Thus, if we set the initial value of this coupling to be smaller than the SM value, we can in principle avoid having a negative Higgs self-coupling just above $T_c$.  However, we still would need a mechanism to restore $y_t$ to its SM value at the weak scale.  Since this is what we have achieved with $\lambda_H$ through the dynamics of $\phi$, we can extend the mechanism to $y_t$.  To do this, we 
introduce a dimension-5 operator 
\beq
O_t = \frac{\phi}{M_t} H^*\epsilon\,{\bar Q_L} t_R + \text{\small H.C.}\,,
\label{Ot}
\eeq                
where $M_t$ is a new UV scale and $\epsilon$ is the two-dimensional Levi-Civita symbol for isospin indices.  In \eq{Ot}, $Q_L$ is the third generation quark doublet and $t_R$ is the right-handed top quark, which we assume has odd $\mathbb Z_2$ parity.  Once $\vev{\phi}\neq 0$, this operator contributes to the low energy Yukawa coupling of the top quark $y^{\rm SM}_t(Q_0)$ which we assume has the SM value, as the notation suggests.  

If we denote the initial value of the top quark Yukawa coupling at $T\gg T_c$ by $y_{0 t}$ we then have 
\beq
y^{\rm SM}_t (Q_0) = y_{0 t}(Q_0) + \frac{\bar \phi}{M_t}\,.
\label{ytSM}
\eeq 
We will choose $y_{0 t} = 0$, so that the Yukawa coupling for the top vanishes before the FOPT when $\vev{\phi}=0$, consistent with the assumed $\mathbb Z_2$ symmetry.  With $y^{\rm SM}_t (Q_0)\approx 0.92$, we find $M_t\approx 1.1 \times 10^3$~TeV.  Hence, before electroweak symmetry breaking, the running of $\lambda_0$ in \eq{running} will be governed by $y_t = y_{0 t}=0$, which is shown in Fig.~\ref{fig:running} as the solid curve.  As one can see, the initial Higgs potential remains stable up to very high scales, beyond the scale $M$ of our effective theory.  In the figure, we have also shown the running of $\lambda_H$ in the SM, given by the dotted curve, which is what we would obtain after electroweak symmetry breaking.

\begin{figure}[t]
\centering
\includegraphics[width=0.8\columnwidth]{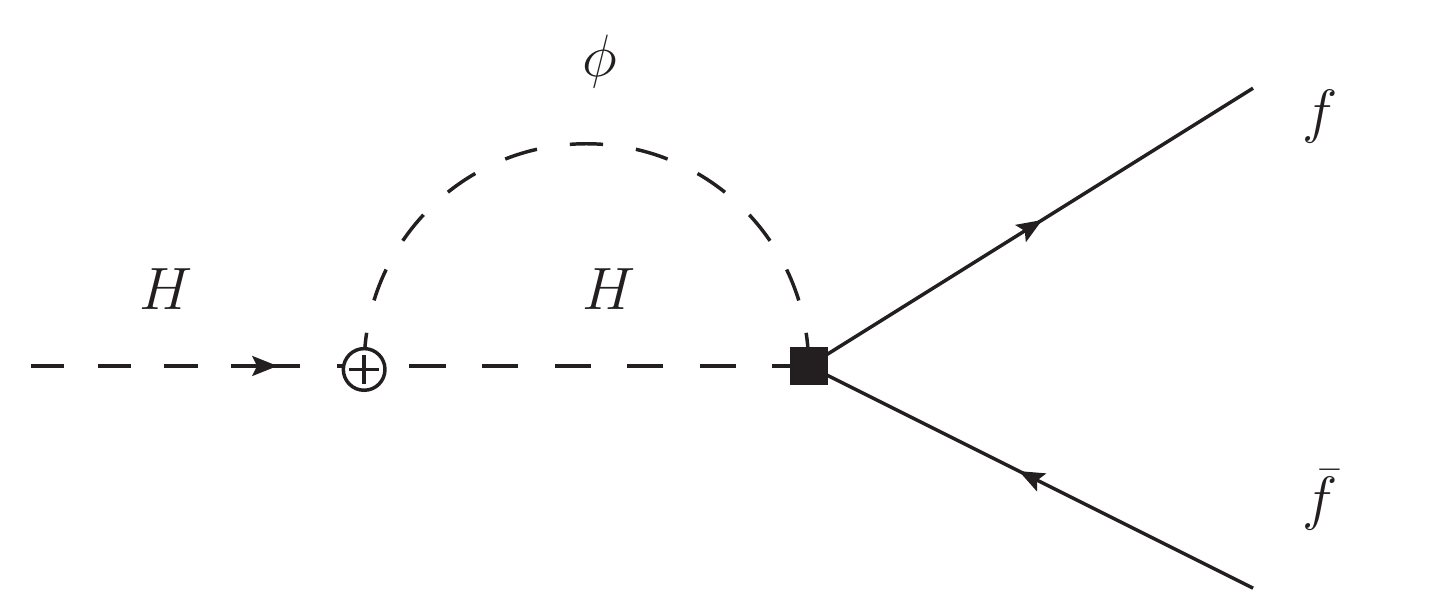}
\caption{One-loop process for generating fermion $f$ Yukawa coupling.  The vertices marked by $\oplus$ and $\blacksquare$ refer to the $\mathbb Z_2$ soft-breaking interaction $\propto \mu$ and a dimension-5 operator of the type in \eq{Of}, respectively.}    
\label{fig:1-loop}
\end{figure}

Note that due to the soft-breaking of the $\mathbb Z_2$, one could in principle generate a Yukawa coupling for the top at 1-loop order, before the FOPT.  However, this contribution is estimated to be 
\beq
\delta y_t(Q_0) \sim \frac{\mu}{16 \pi^2\, M_t} \ln (M_t/Q_0)\,,
\label{delat-yt}
\eeq
which for our reference values is $\ord{10^{-10}}$ and hence completely negligible.  The Feynman diagram of the above 1-loop process, for a fermion $f$, is shown in Fig.\ref{fig:1-loop}. 

\section{Flavor from {\boldmath $\phi$} Dynamics} 

Here, we will examine whether we can generate {\it all} Yukawa couplings $y_f$, where $f$ is a fermion in the SM, from the dynamics of $\phi$.  In the context of our model, the motivation for this would be the $\mathbb Z_2$ symmetry that is so far assumed to be only softly broken.  That is, since we have assumed that $t_R$ is odd under this parity, we cannot write down a generic tree-level Yukawa matrix for quarks without violating the $\mathbb Z_2$.  Also, there is no obvious reason in our model that $t_R$ should be fundamentally different from other spin-1/2 fields.  The extension of the same charge assignment to all fermions in the SM, while not rigorously required, can nonetheless be a natural assumption.  

Let us consider the operator 
\beq
O_f = \frac{\phi}{M_f} H\,{\bar F_L} f_R + \text{\small H.C.}\,,
\label{Of}
\eeq 
where $F$ denotes an $SU(2)_L$ doublet containing the $f$ flavor; this expression is a schematic representation and the proper contraction of gauge indices is assumed depending on the fermion.\footnote{Similar operators have been considered in other contexts, see {\it e.g.} 
Refs.~\cite{Chen:2015vqy,Batell:2017kty}.}  In the above, the scale $M_f$ needs to be chosen so that the correct 
SM Yukawa coupling value is obtained via 
\beq
y^{\rm SM}_f (Q_0) = \frac{\bar \phi}{M_f}\,,
\label{yfSM}
\eeq      
implying 
\beq
M_f = \frac{y^{\rm SM}_t (Q_0)}{y^{\rm SM}_f (Q_0)} M_t.
\label{Mf}
\eeq

The above interaction generates a direct coupling of the fermion $f$ to $\phi$ given by 
\beq
\xi_f = \vev{H}/M_f\,.  
\label{xif}
\eeq
which can potentially lead to phenomenological problems if it is too large.  We will show that if the mixing angle $\theta$ is at an acceptable level, the direct coupling $\xi_f$ is also generically allowed.    To see this, first note that in our model $(\bar\phi/M)^2\sim 0.1$ in order to attain $\lambda_H$ after the FOPT.  Then, \eq{mu-range} implies $\mu\sim 0.1 v^2/{\bar\phi}$ and hence one can deduce $\theta \sim v/{\bar\phi}$.  Using Eqs.~(\ref{yfSM}) and (\ref{xif}), we obtain 
\beq
\theta \,y_f \sim \xi_f\,.
\label{fcouplings}
\eeq
Since we have both mixing-induced and direct coupling for $\phi$, its total coupling to fermions is given by 
\beq
\bar \xi_f \equiv \xi_f + \theta\, y_f^{\rm SM}\,. 
\label{xiebar}
\eeq 
Given the range of allowed values for $\theta$ \cite{Krnjaic:2015mbs}, which covers about an order of magnitude for our reference values of $m_\phi$, $\bar\xi_f$ couplings are also acceptable, since they are within a factor of $\sim 2$ of $\theta y_f$, as implied by \eq{fcouplings}.    

\begin{figure}[t]
\centering
\includegraphics[width=\columnwidth]{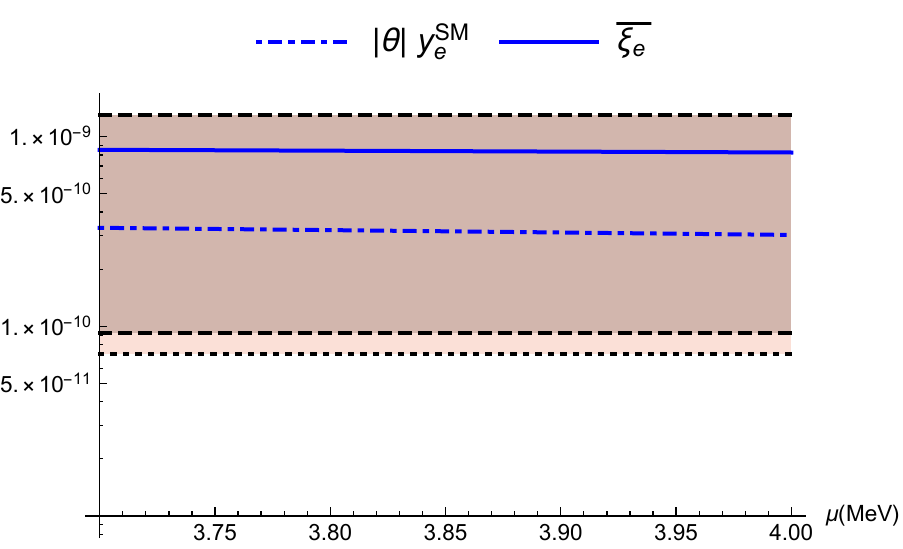}
\caption{Values of $\phi$ coupling to electrons.  The solid line corresponds to the total coupling 
$\bar \xi_e$ and the dot-dashed line is only accounting for $\phi$-Higgs mixing, for $\eps=0.1$ and $M=4\times 10^3$~TeV, as a function of $\mu$ in MeV.  The shaded region between dashed lines approximately corresponds to the allowed values from Ref.~\cite{Krnjaic:2015mbs} and the shaded region extended to the dotted line is set by demanding that $\phi$ decay by BBN \cite{Krnjaic:2015mbs}, based on a revised supernova constraint from Ref.~\cite{Dev:2020eam}.  Here, we have assumed $m_\phi = 7$~MeV, as a typical value from the range (\ref{mphi-range}), for our 
benchmark parameters.}    
\label{fig:phi-e-coupling}
\end{figure}

Using \eq{Mf}, we find $M_e\approx 3.3\times 10^{11}$~GeV.  The interaction (\ref{Of}) generates a direct coupling $\xi_e \phi \,\bar e_L e_R + \text{\small H.C.}$.  Let us choose $m_\phi=7$~MeV, typical of the range (\ref{mphi-range}), as a representative choice, for concreteness and illustrative purposes.  From the results of Ref.~\cite{Krnjaic:2015mbs}, we find that the value of $\theta y_e$  
for $m_\phi \approx7$~MeV must lie approximately in the dark shaded region between the two dashed lines, in Fig.\ref{fig:phi-e-coupling}.  The lower bound is set by supernova physics constraints which were revisited and found to be less severe in Ref.~\cite{Dev:2020eam}.  In that case, the new lower bound moves to the lightly shaded region marked by the dotted line, which is set by requiring a sufficiently short lifetime for $\phi$, {\it i.e.} below $\sim 1$~s, from BBN requirements.  

As we can see from Fig.~\ref{fig:phi-e-coupling}, our benchmark parameters yield acceptable values  for $\theta y_e$.  However, we must also include the effect of $\xi_e$ if we assume the flavor scenario presented above.  
Accounting for both $\phi$-Higgs mixing and $\xi_e>0$, we have plotted the value of $\bar \xi_e$ in Fig.~\ref{fig:phi-e-coupling}, which as one can see lies inside the allowed region of $\phi$-$e$ coupling, in agreement with the preceding general discussion.  Also, note that the overall magnitude of the coupling remains close to that from mixing alone and hence $\phi$ remains long-lived, as deduced 
from \eq{tauphi}, via the replacement $\theta y_e \to \bar \xi_e$.  

We note that the interaction in \eq{Ot}, analogous to the case of the electron, induces a coupling of the top quark to $\phi$ given by 
$\xi_t \phi \,\bar t_L t_R + \text{\small H.C.}$ 
For our benchmark choice of parameters, we find $\xi_t \approx 1.6\times 10^{-4}$.  This is of order a $\phi$-top coupling induced by the $\phi$-Higgs mixing angle $\theta$, because the top Yukawa coupling attains its SM value $y_t^{\rm SM}\approx 1$ after electroweak symmetry breaking in our model.  Assuming no cancellation between the two effects, we find $\bar \xi_t \approx 2.6 \times 10^{-4}$ for  our benchmark values, which can accommodate known constraints \cite{Foroughi-Abari:2020gju}.  

For Dirac neutrinos, we have $y^{\rm SM}_\nu (Q_0) \sim 10^{-12}$, which 
suggests $M_\nu \sim 4\times 10^{18}$~GeV close to, but below, the Planck mass.  We hence see that it is in principle possible that the physics of flavor in the SM is governed by the dynamics of $\phi$, with UV scales ranging from $\ord{10^3}$ TeV all the way up to near Planck mass.  Alternatively, this could signify an exponential hierarchy of flavor dependent couplings between $\phi$ and some UV physics responsible for mediation of the SM flavor structure at scales of $\ord{10^3}$ TeV.    

Before closing the discussion of flavor in our model, we will briefly address how transport of chiral asymmetries invoked in electroweak baryogenesis may be realized, if the SM Yukawa couplings arise only after the FOPT.  Nonetheless, we first emphasize that the viability of our FOPT scenario only requires that the top Yukawa coupling, and perhaps by extension quark Yukawa couplings, remain small until after the transition.  This could still leave lepton Yukawa couplings as a possible avenue for generation of the BAU; see for example Refs.~\cite{Chung:2009cb,deVries:2018tgs,Xie:2020wzn} for work along this direction, in different contexts.  However, even if we adopt the view that all SM fermion Yukawa couplings are generated only after the FOPT, other interactions  may be present, as outlined below.

For example, consider a singlet Dirac fermion $\chi$ that carries a unit of lepton number and whose right (left)-handed component is even (odd) under the assumed $\mathbb Z_2$.  One can write down a Yukawa coupling $\lambda H \bar \chi_R L +$ {\small H.C.} that involves the SM lepton doublet $L$, providing a potential route for generating chiral asymmetries.  We may also include an interaction   
$\lambda' \phi \bar \chi_L \chi_R +$ {\small H.C.} that would allow $\chi$ to become more massive than the Higgs once $\vev{\phi} \neq 0$, to avoid inducing a large deviation, governed by $\lambda$, in the Higgs width.  Note that these couplings would not lead to large masses for the observed neutrinos, even if $\lambda\gsim 0.1$.  The above interactions point to possibilities that may allow our proposed $\phi$-induced SM flavor structure to be consistent with viable electroweak baryogenesis.  A fuller treatment - which is beyond the scope of this work - is required for quantitative conclusions.  


We have collected the benchmark parameters used in the preceding discussions, in Table \ref{table:benchmark}.  

\begin{table}[h]
\centering
\begin{tabular}{c|c||c|c}
$\eps$ & $\kappa$ & $M$ (TeV)& $M_t$ (TeV)  \\ \hline
0.1 & $10^{-4}$  & $4.0\times 10^3$ & $1.1 \times 10^3$ 
\end{tabular}
\caption{Benchmark parameters $\{\eps, \kappa, M, M_t\}$, treated as input choices.}
\label{table:benchmark}
\end{table}

\section{Low Energy Tests and Other Signals}

From \eq{tauphi}, we see that in our scenario $\phi$ will be typically identified as missing energy in laboratory experiments.  This provides a possible avenue for testing our proposal in rare meson decay experiments, such as KOTO.  This experiment aims to measure the branching fraction of the rare kaon decay $K_L \to 
\pi^0 \bar \nu \nu$, which is predicted to be $3.4 \pm 0.6 \times 10^{-11}$ in the SM 
\cite{Cirigliano:2011ny,Egana-Ugrinovic:2019wzj}.  

The KOTO experiment had observed a few events 
\cite{Shinohara:2020brf} that appeared to point to an excess, but at low statistical significance \cite{Egana-Ugrinovic:2019wzj}.  The study in Ref.~\cite{Egana-Ugrinovic:2019wzj} considers an interaction of the type in \eq{Ot} and the results of their analysis suggests that with a $\phi$-Higgs mixing parameter or a $\phi$-top coupling $\sim 2\times 10^{-4}$ and $m_\phi\lsim 50$~MeV, one could obtain a 2$\sigma$ explanation of the seeming KOTO excess.  This results from the coupling of $\phi$ to the top quark in the 1-loop penguin diagram that mediates $K_L\to \pi^0 \phi$, with the `invisible' $\phi$ mimicking the missing energy from $\bar \nu \nu$ (see also \cite{Kitahara:2019lws}).  Hence, parameters quite close to our benchmark region could explain that excess with $\bar \xi_t \sim 2 \times 10^{-4}$, at around 2$\sigma$.  

The above measurement has recently been further studied by the KOTO collaboration, leading to identification of new sources of background \cite{Ahn:2020opg}.  The experiment has now determined that the observed events are statistically consistent with the level of background.  Thus, one may not need to invoke new physics to explain the data.  Nonetheless, the target precision of the KOTO experiment for the branching fraction of 
$K_L\to \pi^0 \bar \nu \nu$, which is at the 10\% level\footnote{See the experimental proposal at https://koto.kek.jp/pub/p14.pdf.}, suggests that it could potentially probe values of 
$\bar \xi_t \lsim 10^{-5}$ in the future, providing an experimental test of our model and its variants over a significant region of parameter space.  We note that in our model the penguin diagrams that mediate $K_L\to \pi^0 \phi$ \cite{Egana-Ugrinovic:2019wzj} are dominated by the top quark, since the couplings $\bar\xi_f$ scale with fermion mass.

Before closing, we would like to point out that it can be typically expected that the FOPT in this scenario would lead to primordial gravitational waves, corresponding to temperatures $\sim T_c$.  This signal may be detectable by future observatories like LISA (see, for example, Ref.~\cite{Caprini:2019egz}).  We leave a quantitative examination of this potential signal of our model to other work.

\section{Summary}

In this work, we have introduced a model for a first order electroweak phase transition which employs a light sub-GeV scalar $\phi$ that is weakly coupled to the Higgs.  Such a phase transition provides a necessary ingredient for a successful baryogenesis mechanism.  The consistency of the model requires new interactions between $\phi$ and the top quark, for stability at scales $\gsim 100$~GeV.  This interaction can be consistent with a softly broken parity of the scalar potential.  We proposed that such interactions could be the source of the entire SM flavor structure, via the dynamics of $\phi$.  The low energy effective theory in our proposal is to a good approximation that of the SM.    
Nonetheless, precision low energy data, such as from measurements of rare kaon decays, can provide tests of our model.  We expect the 
KOTO experiment measurements of 
$K_L\to \pi^0 +$ `missing energy' to have significant reach for the regime of parameters discussed in this work.  General considerations regarding a strong first order phase transition at the weak scale also suggest that primordial gravitational wave signals of this scenario could be detected by a future observatory, such as LISA.  In that case, our model provides an interesting scenario where primordial gravitational wave signals are related to rare phenomena at low energies.  Definite quantitative conclusions on that possibility, however, require further analysis that we leave to other work. 

\begin{acknowledgments}
Work supported by the US Department of Energy under Grant Contract DE-SC0012704.
\end{acknowledgments}

\bibliography{FOEWPT-ref-PRD}

\begin{thebibliography}{41}%
\makeatletter
\providecommand \@ifxundefined [1]{%
 \@ifx{#1\undefined}
}%
\providecommand \@ifnum [1]{%
 \ifnum #1\expandafter \@firstoftwo
 \else \expandafter \@secondoftwo
 \fi
}%
\providecommand \@ifx [1]{%
 \ifx #1\expandafter \@firstoftwo
 \else \expandafter \@secondoftwo
 \fi
}%
\providecommand \natexlab [1]{#1}%
\providecommand \enquote  [1]{``#1''}%
\providecommand \bibnamefont  [1]{#1}%
\providecommand \bibfnamefont [1]{#1}%
\providecommand \citenamefont [1]{#1}%
\providecommand \href@noop [0]{\@secondoftwo}%
\providecommand \href [0]{\begingroup \@sanitize@url \@href}%
\providecommand \@href[1]{\@@startlink{#1}\@@href}%
\providecommand \@@href[1]{\endgroup#1\@@endlink}%
\providecommand \@sanitize@url [0]{\catcode `\\12\catcode `\$12\catcode
  `\&12\catcode `\#12\catcode `\^12\catcode `\_12\catcode `\%12\relax}%
\providecommand \@@startlink[1]{}%
\providecommand \@@endlink[0]{}%
\providecommand \url  [0]{\begingroup\@sanitize@url \@url }%
\providecommand \@url [1]{\endgroup\@href {#1}{\urlprefix }}%
\providecommand \urlprefix  [0]{URL }%
\providecommand \Eprint [0]{\href }%
\providecommand \doibase [0]{https://doi.org/}%
\providecommand \selectlanguage [0]{\@gobble}%
\providecommand \bibinfo  [0]{\@secondoftwo}%
\providecommand \bibfield  [0]{\@secondoftwo}%
\providecommand \translation [1]{[#1]}%
\providecommand \BibitemOpen [0]{}%
\providecommand \bibitemStop [0]{}%
\providecommand \bibitemNoStop [0]{.\EOS\space}%
\providecommand \EOS [0]{\spacefactor3000\relax}%
\providecommand \BibitemShut  [1]{\csname bibitem#1\endcsname}%
\let\auto@bib@innerbib\@empty
\bibitem [{\citenamefont {Tanabashi}\ \emph {et~al.}(2018)\citenamefont
  {Tanabashi} \emph {et~al.}}]{Tanabashi:2018oca}%
  \BibitemOpen
  \bibfield  {author} {\bibinfo {author} {\bibfnamefont {M.}~\bibnamefont
  {Tanabashi}} \emph {et~al.} (\bibinfo {collaboration} {Particle Data
  Group}),\ }\bibfield  {title} {\bibinfo {title} {{Review of Particle
  Physics}},\ }\href {https://doi.org/10.1103/PhysRevD.98.030001} {\bibfield
  {journal} {\bibinfo  {journal} {Phys. Rev. D}\ }\textbf {\bibinfo {volume}
  {98}},\ \bibinfo {pages} {030001} (\bibinfo {year} {2018})}\BibitemShut
  {NoStop}%
\bibitem [{\citenamefont {Sakharov}(1991)}]{Sakharov:1967dj}%
  \BibitemOpen
  \bibfield  {author} {\bibinfo {author} {\bibfnamefont {A.}~\bibnamefont
  {Sakharov}},\ }\bibfield  {title} {\bibinfo {title} {{Violation of CP
  Invariance, C asymmetry, and baryon asymmetry of the universe}},\ }\href
  {https://doi.org/10.1070/PU1991v034n05ABEH002497} {\bibfield  {journal}
  {\bibinfo  {journal} {Sov. Phys. Usp.}\ }\textbf {\bibinfo {volume} {34}},\
  \bibinfo {pages} {392} (\bibinfo {year} {1991})}\BibitemShut {NoStop}%
\bibitem [{\citenamefont {Cohen}\ \emph {et~al.}(1990)\citenamefont {Cohen},
  \citenamefont {Kaplan},\ and\ \citenamefont {Nelson}}]{Cohen:1990py}%
  \BibitemOpen
  \bibfield  {author} {\bibinfo {author} {\bibfnamefont {A.~G.}\ \bibnamefont
  {Cohen}}, \bibinfo {author} {\bibfnamefont {D.~B.}\ \bibnamefont {Kaplan}},\
  and\ \bibinfo {author} {\bibfnamefont {A.~E.}\ \bibnamefont {Nelson}},\
  }\bibfield  {title} {\bibinfo {title} {{WEAK SCALE BARYOGENESIS}},\ }\href
  {https://doi.org/10.1016/0370-2693(90)90690-8} {\bibfield  {journal}
  {\bibinfo  {journal} {Phys. Lett. B}\ }\textbf {\bibinfo {volume} {245}},\
  \bibinfo {pages} {561} (\bibinfo {year} {1990})}\BibitemShut {NoStop}%
\bibitem [{\citenamefont {Quiros}(1999)}]{Quiros:1999jp}%
  \BibitemOpen
  \bibfield  {author} {\bibinfo {author} {\bibfnamefont {M.}~\bibnamefont
  {Quiros}},\ }\bibfield  {title} {\bibinfo {title} {{Finite temperature field
  theory and phase transitions}},\ }in\ \href@noop {} {\emph {\bibinfo
  {booktitle} {{ICTP Summer School in High-Energy Physics and Cosmology}}}}\
  (\bibinfo {year} {1999})\ pp.\ \bibinfo {pages} {187--259},\ \Eprint
  {https://arxiv.org/abs/hep-ph/9901312} {arXiv:hep-ph/9901312} \BibitemShut
  {NoStop}%
\bibitem [{\citenamefont {Carena}\ \emph {et~al.}(2005)\citenamefont {Carena},
  \citenamefont {Megevand}, \citenamefont {Quiros},\ and\ \citenamefont
  {Wagner}}]{Carena:2004ha}%
  \BibitemOpen
  \bibfield  {author} {\bibinfo {author} {\bibfnamefont {M.}~\bibnamefont
  {Carena}}, \bibinfo {author} {\bibfnamefont {A.}~\bibnamefont {Megevand}},
  \bibinfo {author} {\bibfnamefont {M.}~\bibnamefont {Quiros}},\ and\ \bibinfo
  {author} {\bibfnamefont {C.~E.}\ \bibnamefont {Wagner}},\ }\bibfield  {title}
  {\bibinfo {title} {{Electroweak baryogenesis and new TeV fermions}},\ }\href
  {https://doi.org/10.1016/j.nuclphysb.2005.03.025} {\bibfield  {journal}
  {\bibinfo  {journal} {Nucl. Phys. B}\ }\textbf {\bibinfo {volume} {716}},\
  \bibinfo {pages} {319} (\bibinfo {year} {2005})},\ \Eprint
  {https://arxiv.org/abs/hep-ph/0410352} {arXiv:hep-ph/0410352} \BibitemShut
  {NoStop}%
\bibitem [{\citenamefont {Nelson}\ \emph {et~al.}(1992)\citenamefont {Nelson},
  \citenamefont {Kaplan},\ and\ \citenamefont {Cohen}}]{Nelson:1991ab}%
  \BibitemOpen
  \bibfield  {author} {\bibinfo {author} {\bibfnamefont {A.}~\bibnamefont
  {Nelson}}, \bibinfo {author} {\bibfnamefont {D.}~\bibnamefont {Kaplan}},\
  and\ \bibinfo {author} {\bibfnamefont {A.~G.}\ \bibnamefont {Cohen}},\
  }\bibfield  {title} {\bibinfo {title} {{Why there is something rather than
  nothing: Matter from weak interactions}},\ }\href
  {https://doi.org/10.1016/0550-3213(92)90440-M} {\bibfield  {journal}
  {\bibinfo  {journal} {Nucl. Phys. B}\ }\textbf {\bibinfo {volume} {373}},\
  \bibinfo {pages} {453} (\bibinfo {year} {1992})}\BibitemShut {NoStop}%
\bibitem [{\citenamefont {Espinosa}\ \emph {et~al.}(1993)\citenamefont
  {Espinosa}, \citenamefont {Quiros},\ and\ \citenamefont
  {Zwirner}}]{Espinosa:1993yi}%
  \BibitemOpen
  \bibfield  {author} {\bibinfo {author} {\bibfnamefont {J.}~\bibnamefont
  {Espinosa}}, \bibinfo {author} {\bibfnamefont {M.}~\bibnamefont {Quiros}},\
  and\ \bibinfo {author} {\bibfnamefont {F.}~\bibnamefont {Zwirner}},\
  }\bibfield  {title} {\bibinfo {title} {{On the electroweak phase transition
  in the minimal supersymmetric Standard Model}},\ }\href
  {https://doi.org/10.1016/0370-2693(93)90199-R} {\bibfield  {journal}
  {\bibinfo  {journal} {Phys. Lett. B}\ }\textbf {\bibinfo {volume} {307}},\
  \bibinfo {pages} {106} (\bibinfo {year} {1993})},\ \Eprint
  {https://arxiv.org/abs/hep-ph/9303317} {arXiv:hep-ph/9303317} \BibitemShut
  {NoStop}%
\bibitem [{\citenamefont {Carena}\ \emph {et~al.}(1996)\citenamefont {Carena},
  \citenamefont {Quiros},\ and\ \citenamefont {Wagner}}]{Carena:1996wj}%
  \BibitemOpen
  \bibfield  {author} {\bibinfo {author} {\bibfnamefont {M.}~\bibnamefont
  {Carena}}, \bibinfo {author} {\bibfnamefont {M.}~\bibnamefont {Quiros}},\
  and\ \bibinfo {author} {\bibfnamefont {C.}~\bibnamefont {Wagner}},\
  }\bibfield  {title} {\bibinfo {title} {{Opening the window for electroweak
  baryogenesis}},\ }\href {https://doi.org/10.1016/0370-2693(96)00475-3}
  {\bibfield  {journal} {\bibinfo  {journal} {Phys. Lett. B}\ }\textbf
  {\bibinfo {volume} {380}},\ \bibinfo {pages} {81} (\bibinfo {year} {1996})},\
  \Eprint {https://arxiv.org/abs/hep-ph/9603420} {arXiv:hep-ph/9603420}
  \BibitemShut {NoStop}%
\bibitem [{\citenamefont {Kanemura}\ \emph {et~al.}(2005)\citenamefont
  {Kanemura}, \citenamefont {Okada},\ and\ \citenamefont
  {Senaha}}]{Kanemura:2004ch}%
  \BibitemOpen
  \bibfield  {author} {\bibinfo {author} {\bibfnamefont {S.}~\bibnamefont
  {Kanemura}}, \bibinfo {author} {\bibfnamefont {Y.}~\bibnamefont {Okada}},\
  and\ \bibinfo {author} {\bibfnamefont {E.}~\bibnamefont {Senaha}},\
  }\bibfield  {title} {\bibinfo {title} {{Electroweak baryogenesis and quantum
  corrections to the triple Higgs boson coupling}},\ }\href
  {https://doi.org/10.1016/j.physletb.2004.12.004} {\bibfield  {journal}
  {\bibinfo  {journal} {Phys. Lett. B}\ }\textbf {\bibinfo {volume} {606}},\
  \bibinfo {pages} {361} (\bibinfo {year} {2005})},\ \Eprint
  {https://arxiv.org/abs/hep-ph/0411354} {arXiv:hep-ph/0411354} \BibitemShut
  {NoStop}%
\bibitem [{\citenamefont {Profumo}\ \emph {et~al.}(2007)\citenamefont
  {Profumo}, \citenamefont {Ramsey-Musolf},\ and\ \citenamefont
  {Shaughnessy}}]{Profumo:2007wc}%
  \BibitemOpen
  \bibfield  {author} {\bibinfo {author} {\bibfnamefont {S.}~\bibnamefont
  {Profumo}}, \bibinfo {author} {\bibfnamefont {M.~J.}\ \bibnamefont
  {Ramsey-Musolf}},\ and\ \bibinfo {author} {\bibfnamefont {G.}~\bibnamefont
  {Shaughnessy}},\ }\bibfield  {title} {\bibinfo {title} {{Singlet Higgs
  phenomenology and the electroweak phase transition}},\ }\href
  {https://doi.org/10.1088/1126-6708/2007/08/010} {\bibfield  {journal}
  {\bibinfo  {journal} {JHEP}\ }\textbf {\bibinfo {volume} {08}},\ \bibinfo
  {pages} {010}},\ \Eprint {https://arxiv.org/abs/0705.2425} {arXiv:0705.2425
  [hep-ph]} \BibitemShut {NoStop}%
\bibitem [{\citenamefont {Noble}\ and\ \citenamefont
  {Perelstein}(2008)}]{Noble:2007kk}%
  \BibitemOpen
  \bibfield  {author} {\bibinfo {author} {\bibfnamefont {A.}~\bibnamefont
  {Noble}}\ and\ \bibinfo {author} {\bibfnamefont {M.}~\bibnamefont
  {Perelstein}},\ }\bibfield  {title} {\bibinfo {title} {{Higgs self-coupling
  as a probe of electroweak phase transition}},\ }\href
  {https://doi.org/10.1103/PhysRevD.78.063518} {\bibfield  {journal} {\bibinfo
  {journal} {Phys. Rev. D}\ }\textbf {\bibinfo {volume} {78}},\ \bibinfo
  {pages} {063518} (\bibinfo {year} {2008})},\ \Eprint
  {https://arxiv.org/abs/0711.3018} {arXiv:0711.3018 [hep-ph]} \BibitemShut
  {NoStop}%
\bibitem [{\citenamefont {No}\ and\ \citenamefont
  {Ramsey-Musolf}(2014)}]{No:2013wsa}%
  \BibitemOpen
  \bibfield  {author} {\bibinfo {author} {\bibfnamefont {J.~M.}\ \bibnamefont
  {No}}\ and\ \bibinfo {author} {\bibfnamefont {M.}~\bibnamefont
  {Ramsey-Musolf}},\ }\bibfield  {title} {\bibinfo {title} {{Probing the Higgs
  Portal at the LHC Through Resonant di-Higgs Production}},\ }\href
  {https://doi.org/10.1103/PhysRevD.89.095031} {\bibfield  {journal} {\bibinfo
  {journal} {Phys. Rev. D}\ }\textbf {\bibinfo {volume} {89}},\ \bibinfo
  {pages} {095031} (\bibinfo {year} {2014})},\ \Eprint
  {https://arxiv.org/abs/1310.6035} {arXiv:1310.6035 [hep-ph]} \BibitemShut
  {NoStop}%
\bibitem [{\citenamefont {Katz}\ and\ \citenamefont
  {Perelstein}(2014)}]{Katz:2014bha}%
  \BibitemOpen
  \bibfield  {author} {\bibinfo {author} {\bibfnamefont {A.}~\bibnamefont
  {Katz}}\ and\ \bibinfo {author} {\bibfnamefont {M.}~\bibnamefont
  {Perelstein}},\ }\bibfield  {title} {\bibinfo {title} {{Higgs Couplings and
  Electroweak Phase Transition}},\ }\href
  {https://doi.org/10.1007/JHEP07(2014)108} {\bibfield  {journal} {\bibinfo
  {journal} {JHEP}\ }\textbf {\bibinfo {volume} {07}},\ \bibinfo {pages}
  {108}},\ \Eprint {https://arxiv.org/abs/1401.1827} {arXiv:1401.1827 [hep-ph]}
  \BibitemShut {NoStop}%
\bibitem [{\citenamefont {Basler}\ \emph {et~al.}(2017)\citenamefont {Basler},
  \citenamefont {Krause}, \citenamefont {Muhlleitner}, \citenamefont
  {Wittbrodt},\ and\ \citenamefont {Wlotzka}}]{Basler:2016obg}%
  \BibitemOpen
  \bibfield  {author} {\bibinfo {author} {\bibfnamefont {P.}~\bibnamefont
  {Basler}}, \bibinfo {author} {\bibfnamefont {M.}~\bibnamefont {Krause}},
  \bibinfo {author} {\bibfnamefont {M.}~\bibnamefont {Muhlleitner}}, \bibinfo
  {author} {\bibfnamefont {J.}~\bibnamefont {Wittbrodt}},\ and\ \bibinfo
  {author} {\bibfnamefont {A.}~\bibnamefont {Wlotzka}},\ }\bibfield  {title}
  {\bibinfo {title} {{Strong First Order Electroweak Phase Transition in the
  CP-Conserving 2HDM Revisited}},\ }\href
  {https://doi.org/10.1007/JHEP02(2017)121} {\bibfield  {journal} {\bibinfo
  {journal} {JHEP}\ }\textbf {\bibinfo {volume} {02}},\ \bibinfo {pages}
  {121}},\ \Eprint {https://arxiv.org/abs/1612.04086} {arXiv:1612.04086
  [hep-ph]} \BibitemShut {NoStop}%
\bibitem [{\citenamefont {Kurup}\ and\ \citenamefont
  {Perelstein}(2017)}]{Kurup:2017dzf}%
  \BibitemOpen
  \bibfield  {author} {\bibinfo {author} {\bibfnamefont {G.}~\bibnamefont
  {Kurup}}\ and\ \bibinfo {author} {\bibfnamefont {M.}~\bibnamefont
  {Perelstein}},\ }\bibfield  {title} {\bibinfo {title} {{Dynamics of
  Electroweak Phase Transition In Singlet-Scalar Extension of the Standard
  Model}},\ }\href {https://doi.org/10.1103/PhysRevD.96.015036} {\bibfield
  {journal} {\bibinfo  {journal} {Phys. Rev. D}\ }\textbf {\bibinfo {volume}
  {96}},\ \bibinfo {pages} {015036} (\bibinfo {year} {2017})},\ \Eprint
  {https://arxiv.org/abs/1704.03381} {arXiv:1704.03381 [hep-ph]} \BibitemShut
  {NoStop}%
\bibitem [{\citenamefont {Chen}\ \emph {et~al.}(2017)\citenamefont {Chen},
  \citenamefont {Kozaczuk},\ and\ \citenamefont {Lewis}}]{Chen:2017qcz}%
  \BibitemOpen
  \bibfield  {author} {\bibinfo {author} {\bibfnamefont {C.-Y.}\ \bibnamefont
  {Chen}}, \bibinfo {author} {\bibfnamefont {J.}~\bibnamefont {Kozaczuk}},\
  and\ \bibinfo {author} {\bibfnamefont {I.~M.}\ \bibnamefont {Lewis}},\
  }\bibfield  {title} {\bibinfo {title} {{Non-resonant Collider Signatures of a
  Singlet-Driven Electroweak Phase Transition}},\ }\href
  {https://doi.org/10.1007/JHEP08(2017)096} {\bibfield  {journal} {\bibinfo
  {journal} {JHEP}\ }\textbf {\bibinfo {volume} {08}},\ \bibinfo {pages}
  {096}},\ \Eprint {https://arxiv.org/abs/1704.05844} {arXiv:1704.05844
  [hep-ph]} \BibitemShut {NoStop}%
\bibitem [{\citenamefont {Ramsey-Musolf}(2020)}]{Ramsey-Musolf:2019lsf}%
  \BibitemOpen
  \bibfield  {author} {\bibinfo {author} {\bibfnamefont {M.~J.}\ \bibnamefont
  {Ramsey-Musolf}},\ }\bibfield  {title} {\bibinfo {title} {{The electroweak
  phase transition: a collider target}},\ }\href
  {https://doi.org/10.1007/JHEP09(2020)179} {\bibfield  {journal} {\bibinfo
  {journal} {JHEP}\ }\textbf {\bibinfo {volume} {09}},\ \bibinfo {pages}
  {179}},\ \Eprint {https://arxiv.org/abs/1912.07189} {arXiv:1912.07189
  [hep-ph]} \BibitemShut {NoStop}%
\bibitem [{\citenamefont {Basler}\ \emph {et~al.}(2020)\citenamefont {Basler},
  \citenamefont {M\"uhlleitner},\ and\ \citenamefont
  {M\"uller}}]{Basler:2019iuu}%
  \BibitemOpen
  \bibfield  {author} {\bibinfo {author} {\bibfnamefont {P.}~\bibnamefont
  {Basler}}, \bibinfo {author} {\bibfnamefont {M.}~\bibnamefont
  {M\"uhlleitner}},\ and\ \bibinfo {author} {\bibfnamefont {J.}~\bibnamefont
  {M\"uller}},\ }\bibfield  {title} {\bibinfo {title} {{Electroweak Phase
  Transition in Non-Minimal Higgs Sectors}},\ }\href
  {https://doi.org/10.1007/JHEP05(2020)016} {\bibfield  {journal} {\bibinfo
  {journal} {JHEP}\ }\textbf {\bibinfo {volume} {05}},\ \bibinfo {pages}
  {016}},\ \Eprint {https://arxiv.org/abs/1912.10477} {arXiv:1912.10477
  [hep-ph]} \BibitemShut {NoStop}%
\bibitem [{\citenamefont {Jeong}\ \emph {et~al.}(2019)\citenamefont {Jeong},
  \citenamefont {Jung},\ and\ \citenamefont {Shin}}]{Jeong:2018ucz}%
  \BibitemOpen
  \bibfield  {author} {\bibinfo {author} {\bibfnamefont {K.~S.}\ \bibnamefont
  {Jeong}}, \bibinfo {author} {\bibfnamefont {T.~H.}\ \bibnamefont {Jung}},\
  and\ \bibinfo {author} {\bibfnamefont {C.~S.}\ \bibnamefont {Shin}},\
  }\bibfield  {title} {\bibinfo {title} {{Axionic Electroweak Baryogenesis}},\
  }\href {https://doi.org/10.1016/j.physletb.2019.01.036} {\bibfield  {journal}
  {\bibinfo  {journal} {Phys. Lett. B}\ }\textbf {\bibinfo {volume} {790}},\
  \bibinfo {pages} {326} (\bibinfo {year} {2019})},\ \Eprint
  {https://arxiv.org/abs/1806.02591} {arXiv:1806.02591 [hep-ph]} \BibitemShut
  {NoStop}%
\bibitem [{\citenamefont {Kozaczuk}\ \emph {et~al.}(2020)\citenamefont
  {Kozaczuk}, \citenamefont {Ramsey-Musolf},\ and\ \citenamefont
  {Shelton}}]{Kozaczuk:2019pet}%
  \BibitemOpen
  \bibfield  {author} {\bibinfo {author} {\bibfnamefont {J.}~\bibnamefont
  {Kozaczuk}}, \bibinfo {author} {\bibfnamefont {M.~J.}\ \bibnamefont
  {Ramsey-Musolf}},\ and\ \bibinfo {author} {\bibfnamefont {J.}~\bibnamefont
  {Shelton}},\ }\bibfield  {title} {\bibinfo {title} {{Exotic Higgs boson
  decays and the electroweak phase transition}},\ }\href
  {https://doi.org/10.1103/PhysRevD.101.115035} {\bibfield  {journal} {\bibinfo
   {journal} {Phys. Rev. D}\ }\textbf {\bibinfo {volume} {101}},\ \bibinfo
  {pages} {115035} (\bibinfo {year} {2020})},\ \Eprint
  {https://arxiv.org/abs/1911.10210} {arXiv:1911.10210 [hep-ph]} \BibitemShut
  {NoStop}%
\bibitem [{\citenamefont {Jeong}\ \emph {et~al.}(2020)\citenamefont {Jeong},
  \citenamefont {Jung},\ and\ \citenamefont {Shin}}]{Jeong:2018jqe}%
  \BibitemOpen
  \bibfield  {author} {\bibinfo {author} {\bibfnamefont {K.~S.}\ \bibnamefont
  {Jeong}}, \bibinfo {author} {\bibfnamefont {T.~H.}\ \bibnamefont {Jung}},\
  and\ \bibinfo {author} {\bibfnamefont {C.~S.}\ \bibnamefont {Shin}},\
  }\bibfield  {title} {\bibinfo {title} {{Adiabatic electroweak baryogenesis
  driven by an axionlike particle}},\ }\href
  {https://doi.org/10.1103/PhysRevD.101.035009} {\bibfield  {journal} {\bibinfo
   {journal} {Phys. Rev. D}\ }\textbf {\bibinfo {volume} {101}},\ \bibinfo
  {pages} {035009} (\bibinfo {year} {2020})},\ \Eprint
  {https://arxiv.org/abs/1811.03294} {arXiv:1811.03294 [hep-ph]} \BibitemShut
  {NoStop}%
\bibitem [{\citenamefont {Ghosh}\ \emph {et~al.}()\citenamefont {Ghosh},
  \citenamefont {Guo}, \citenamefont {Han},\ and\ \citenamefont
  {Liu}}]{Ghosh:2020ipy}%
  \BibitemOpen
  \bibfield  {author} {\bibinfo {author} {\bibfnamefont {T.}~\bibnamefont
  {Ghosh}}, \bibinfo {author} {\bibfnamefont {H.-K.}\ \bibnamefont {Guo}},
  \bibinfo {author} {\bibfnamefont {T.}~\bibnamefont {Han}},\ and\ \bibinfo
  {author} {\bibfnamefont {H.}~\bibnamefont {Liu}},\ }\bibfield  {title}
  {\bibinfo {title} {{Electroweak Phase Transition with an SU(2) Dark
  Sector}},\ }\href@noop {} {\ }\Eprint {https://arxiv.org/abs/2012.09758}
  {arXiv:2012.09758 [hep-ph]} \BibitemShut {NoStop}%
\bibitem [{\citenamefont {Grojean}\ \emph {et~al.}(2005)\citenamefont
  {Grojean}, \citenamefont {Servant},\ and\ \citenamefont
  {Wells}}]{Grojean:2004xa}%
  \BibitemOpen
  \bibfield  {author} {\bibinfo {author} {\bibfnamefont {C.}~\bibnamefont
  {Grojean}}, \bibinfo {author} {\bibfnamefont {G.}~\bibnamefont {Servant}},\
  and\ \bibinfo {author} {\bibfnamefont {J.~D.}\ \bibnamefont {Wells}},\
  }\bibfield  {title} {\bibinfo {title} {{First-order electroweak phase
  transition in the standard model with a low cutoff}},\ }\href
  {https://doi.org/10.1103/PhysRevD.71.036001} {\bibfield  {journal} {\bibinfo
  {journal} {Phys. Rev. D}\ }\textbf {\bibinfo {volume} {71}},\ \bibinfo
  {pages} {036001} (\bibinfo {year} {2005})},\ \Eprint
  {https://arxiv.org/abs/hep-ph/0407019} {arXiv:hep-ph/0407019} \BibitemShut
  {NoStop}%
\bibitem [{\citenamefont {Delaunay}\ \emph {et~al.}(2008)\citenamefont
  {Delaunay}, \citenamefont {Grojean},\ and\ \citenamefont
  {Wells}}]{Delaunay:2007wb}%
  \BibitemOpen
  \bibfield  {author} {\bibinfo {author} {\bibfnamefont {C.}~\bibnamefont
  {Delaunay}}, \bibinfo {author} {\bibfnamefont {C.}~\bibnamefont {Grojean}},\
  and\ \bibinfo {author} {\bibfnamefont {J.~D.}\ \bibnamefont {Wells}},\
  }\bibfield  {title} {\bibinfo {title} {{Dynamics of Non-renormalizable
  Electroweak Symmetry Breaking}},\ }\href
  {https://doi.org/10.1088/1126-6708/2008/04/029} {\bibfield  {journal}
  {\bibinfo  {journal} {JHEP}\ }\textbf {\bibinfo {volume} {04}},\ \bibinfo
  {pages} {029}},\ \Eprint {https://arxiv.org/abs/0711.2511} {arXiv:0711.2511
  [hep-ph]} \BibitemShut {NoStop}%
\bibitem [{\citenamefont {Davoudiasl}\ \emph {et~al.}(2013)\citenamefont
  {Davoudiasl}, \citenamefont {Lewis},\ and\ \citenamefont
  {Ponton}}]{Davoudiasl:2012tu}%
  \BibitemOpen
  \bibfield  {author} {\bibinfo {author} {\bibfnamefont {H.}~\bibnamefont
  {Davoudiasl}}, \bibinfo {author} {\bibfnamefont {I.}~\bibnamefont {Lewis}},\
  and\ \bibinfo {author} {\bibfnamefont {E.}~\bibnamefont {Ponton}},\
  }\bibfield  {title} {\bibinfo {title} {{Electroweak Phase Transition, Higgs
  Diphoton Rate, and New Heavy Fermions}},\ }\href
  {https://doi.org/10.1103/PhysRevD.87.093001} {\bibfield  {journal} {\bibinfo
  {journal} {Phys. Rev. D}\ }\textbf {\bibinfo {volume} {87}},\ \bibinfo
  {pages} {093001} (\bibinfo {year} {2013})},\ \Eprint
  {https://arxiv.org/abs/1211.3449} {arXiv:1211.3449 [hep-ph]} \BibitemShut
  {NoStop}%
\bibitem [{\citenamefont {Baldes}\ \emph {et~al.}(2016)\citenamefont {Baldes},
  \citenamefont {Konstandin},\ and\ \citenamefont {Servant}}]{Baldes:2016gaf}%
  \BibitemOpen
  \bibfield  {author} {\bibinfo {author} {\bibfnamefont {I.}~\bibnamefont
  {Baldes}}, \bibinfo {author} {\bibfnamefont {T.}~\bibnamefont {Konstandin}},\
  and\ \bibinfo {author} {\bibfnamefont {G.}~\bibnamefont {Servant}},\
  }\bibfield  {title} {\bibinfo {title} {{Flavor Cosmology: Dynamical Yukawas
  in the Froggatt-Nielsen Mechanism}},\ }\href
  {https://doi.org/10.1007/JHEP12(2016)073} {\bibfield  {journal} {\bibinfo
  {journal} {JHEP}\ }\textbf {\bibinfo {volume} {12}},\ \bibinfo {pages}
  {073}},\ \Eprint {https://arxiv.org/abs/1608.03254} {arXiv:1608.03254
  [hep-ph]} \BibitemShut {NoStop}%
\bibitem [{\citenamefont {Batell}\ \emph {et~al.}(2013)\citenamefont {Batell},
  \citenamefont {Jung},\ and\ \citenamefont {Lee}}]{Batell:2012zw}%
  \BibitemOpen
  \bibfield  {author} {\bibinfo {author} {\bibfnamefont {B.}~\bibnamefont
  {Batell}}, \bibinfo {author} {\bibfnamefont {S.}~\bibnamefont {Jung}},\ and\
  \bibinfo {author} {\bibfnamefont {H.~M.}\ \bibnamefont {Lee}},\ }\bibfield
  {title} {\bibinfo {title} {{Singlet Assisted Vacuum Stability and the Higgs
  to Diphoton Rate}},\ }\href {https://doi.org/10.1007/JHEP01(2013)135}
  {\bibfield  {journal} {\bibinfo  {journal} {JHEP}\ }\textbf {\bibinfo
  {volume} {01}},\ \bibinfo {pages} {135}},\ \Eprint
  {https://arxiv.org/abs/1211.2449} {arXiv:1211.2449 [hep-ph]} \BibitemShut
  {NoStop}%
\bibitem [{\citenamefont {Chen}\ \emph {et~al.}(2016)\citenamefont {Chen},
  \citenamefont {Davoudiasl}, \citenamefont {Marciano},\ and\ \citenamefont
  {Zhang}}]{Chen:2015vqy}%
  \BibitemOpen
  \bibfield  {author} {\bibinfo {author} {\bibfnamefont {C.-Y.}\ \bibnamefont
  {Chen}}, \bibinfo {author} {\bibfnamefont {H.}~\bibnamefont {Davoudiasl}},
  \bibinfo {author} {\bibfnamefont {W.~J.}\ \bibnamefont {Marciano}},\ and\
  \bibinfo {author} {\bibfnamefont {C.}~\bibnamefont {Zhang}},\ }\bibfield
  {title} {\bibinfo {title} {{Implications of a light \textquotedblleft{}dark
  Higgs\textquotedblright{} solution to the $g_\mu$-2 discrepancy}},\ }\href
  {https://doi.org/10.1103/PhysRevD.93.035006} {\bibfield  {journal} {\bibinfo
  {journal} {Phys. Rev. D}\ }\textbf {\bibinfo {volume} {93}},\ \bibinfo
  {pages} {035006} (\bibinfo {year} {2016})},\ \Eprint
  {https://arxiv.org/abs/1511.04715} {arXiv:1511.04715 [hep-ph]} \BibitemShut
  {NoStop}%
\bibitem [{\citenamefont {Batell}\ \emph {et~al.}(2018)\citenamefont {Batell},
  \citenamefont {Freitas}, \citenamefont {Ismail},\ and\ \citenamefont
  {Mckeen}}]{Batell:2017kty}%
  \BibitemOpen
  \bibfield  {author} {\bibinfo {author} {\bibfnamefont {B.}~\bibnamefont
  {Batell}}, \bibinfo {author} {\bibfnamefont {A.}~\bibnamefont {Freitas}},
  \bibinfo {author} {\bibfnamefont {A.}~\bibnamefont {Ismail}},\ and\ \bibinfo
  {author} {\bibfnamefont {D.}~\bibnamefont {Mckeen}},\ }\bibfield  {title}
  {\bibinfo {title} {{Flavor-specific scalar mediators}},\ }\href
  {https://doi.org/10.1103/PhysRevD.98.055026} {\bibfield  {journal} {\bibinfo
  {journal} {Phys. Rev. D}\ }\textbf {\bibinfo {volume} {98}},\ \bibinfo
  {pages} {055026} (\bibinfo {year} {2018})},\ \Eprint
  {https://arxiv.org/abs/1712.10022} {arXiv:1712.10022 [hep-ph]} \BibitemShut
  {NoStop}%
\bibitem [{\citenamefont {Krnjaic}(2016)}]{Krnjaic:2015mbs}%
  \BibitemOpen
  \bibfield  {author} {\bibinfo {author} {\bibfnamefont {G.}~\bibnamefont
  {Krnjaic}},\ }\bibfield  {title} {\bibinfo {title} {{Probing Light Thermal
  Dark-Matter With a Higgs Portal Mediator}},\ }\href
  {https://doi.org/10.1103/PhysRevD.94.073009} {\bibfield  {journal} {\bibinfo
  {journal} {Phys. Rev. D}\ }\textbf {\bibinfo {volume} {94}},\ \bibinfo
  {pages} {073009} (\bibinfo {year} {2016})},\ \Eprint
  {https://arxiv.org/abs/1512.04119} {arXiv:1512.04119 [hep-ph]} \BibitemShut
  {NoStop}%
\bibitem [{\citenamefont {Dev}\ \emph {et~al.}(2020)\citenamefont {Dev},
  \citenamefont {Mohapatra},\ and\ \citenamefont {Zhang}}]{Dev:2020eam}%
  \BibitemOpen
  \bibfield  {author} {\bibinfo {author} {\bibfnamefont {P.~B.}\ \bibnamefont
  {Dev}}, \bibinfo {author} {\bibfnamefont {R.~N.}\ \bibnamefont {Mohapatra}},\
  and\ \bibinfo {author} {\bibfnamefont {Y.}~\bibnamefont {Zhang}},\ }\bibfield
   {title} {\bibinfo {title} {{Revisiting supernova constraints on a light
  CP-even scalar}},\ }\href {https://doi.org/10.1088/1475-7516/2020/08/003}
  {\bibfield  {journal} {\bibinfo  {journal} {JCAP}\ }\textbf {\bibinfo
  {volume} {08}},\ \bibinfo {pages} {003}},\ \bibinfo {note} {[Erratum: JCAP
  11, E01 (2020)]},\ \Eprint {https://arxiv.org/abs/2005.00490}
  {arXiv:2005.00490 [hep-ph]} \BibitemShut {NoStop}%
\bibitem [{\citenamefont {Foroughi-Abari}\ and\ \citenamefont
  {Ritz}(2020)}]{Foroughi-Abari:2020gju}%
  \BibitemOpen
  \bibfield  {author} {\bibinfo {author} {\bibfnamefont {S.}~\bibnamefont
  {Foroughi-Abari}}\ and\ \bibinfo {author} {\bibfnamefont {A.}~\bibnamefont
  {Ritz}},\ }\bibfield  {title} {\bibinfo {title} {{LSND Constraints on the
  Higgs Portal}},\ }\href {https://doi.org/10.1103/PhysRevD.102.035015}
  {\bibfield  {journal} {\bibinfo  {journal} {Phys. Rev. D}\ }\textbf {\bibinfo
  {volume} {102}},\ \bibinfo {pages} {035015} (\bibinfo {year} {2020})},\
  \Eprint {https://arxiv.org/abs/2004.14515} {arXiv:2004.14515 [hep-ph]}
  \BibitemShut {NoStop}%
\bibitem [{\citenamefont {Chung}\ \emph {et~al.}(2010)\citenamefont {Chung},
  \citenamefont {Garbrecht}, \citenamefont {Ramsey-Musolf},\ and\ \citenamefont
  {Tulin}}]{Chung:2009cb}%
  \BibitemOpen
  \bibfield  {author} {\bibinfo {author} {\bibfnamefont {D.~J.~H.}\
  \bibnamefont {Chung}}, \bibinfo {author} {\bibfnamefont {B.}~\bibnamefont
  {Garbrecht}}, \bibinfo {author} {\bibfnamefont {M.~J.}\ \bibnamefont
  {Ramsey-Musolf}},\ and\ \bibinfo {author} {\bibfnamefont {S.}~\bibnamefont
  {Tulin}},\ }\bibfield  {title} {\bibinfo {title} {{Lepton-mediated
  electroweak baryogenesis}},\ }\href
  {https://doi.org/10.1103/PhysRevD.81.063506} {\bibfield  {journal} {\bibinfo
  {journal} {Phys. Rev. D}\ }\textbf {\bibinfo {volume} {81}},\ \bibinfo
  {pages} {063506} (\bibinfo {year} {2010})},\ \Eprint
  {https://arxiv.org/abs/0905.4509} {arXiv:0905.4509 [hep-ph]} \BibitemShut
  {NoStop}%
\bibitem [{\citenamefont {De~Vries}\ \emph {et~al.}(2019)\citenamefont
  {De~Vries}, \citenamefont {Postma},\ and\ \citenamefont {van~de
  Vis}}]{deVries:2018tgs}%
  \BibitemOpen
  \bibfield  {author} {\bibinfo {author} {\bibfnamefont {J.}~\bibnamefont
  {De~Vries}}, \bibinfo {author} {\bibfnamefont {M.}~\bibnamefont {Postma}},\
  and\ \bibinfo {author} {\bibfnamefont {J.}~\bibnamefont {van~de Vis}},\
  }\bibfield  {title} {\bibinfo {title} {{The role of leptons in electroweak
  baryogenesis}},\ }\href {https://doi.org/10.1007/JHEP04(2019)024} {\bibfield
  {journal} {\bibinfo  {journal} {JHEP}\ }\textbf {\bibinfo {volume} {04}},\
  \bibinfo {pages} {024}},\ \Eprint {https://arxiv.org/abs/1811.11104}
  {arXiv:1811.11104 [hep-ph]} \BibitemShut {NoStop}%
\bibitem [{\citenamefont {Xie}(2021)}]{Xie:2020wzn}%
  \BibitemOpen
  \bibfield  {author} {\bibinfo {author} {\bibfnamefont {K.-P.}\ \bibnamefont
  {Xie}},\ }\bibfield  {title} {\bibinfo {title} {{Lepton-mediated electroweak
  baryogenesis, gravitational waves and the $4\tau$ final state at the
  collider}},\ }\href {https://doi.org/10.1007/JHEP02(2021)090} {\bibfield
  {journal} {\bibinfo  {journal} {JHEP}\ }\textbf {\bibinfo {volume} {02}},\
  \bibinfo {pages} {090}},\ \Eprint {https://arxiv.org/abs/2011.04821}
  {arXiv:2011.04821 [hep-ph]} \BibitemShut {NoStop}%
\bibitem [{\citenamefont {Cirigliano}\ \emph {et~al.}(2012)\citenamefont
  {Cirigliano}, \citenamefont {Ecker}, \citenamefont {Neufeld}, \citenamefont
  {Pich},\ and\ \citenamefont {Portoles}}]{Cirigliano:2011ny}%
  \BibitemOpen
  \bibfield  {author} {\bibinfo {author} {\bibfnamefont {V.}~\bibnamefont
  {Cirigliano}}, \bibinfo {author} {\bibfnamefont {G.}~\bibnamefont {Ecker}},
  \bibinfo {author} {\bibfnamefont {H.}~\bibnamefont {Neufeld}}, \bibinfo
  {author} {\bibfnamefont {A.}~\bibnamefont {Pich}},\ and\ \bibinfo {author}
  {\bibfnamefont {J.}~\bibnamefont {Portoles}},\ }\bibfield  {title} {\bibinfo
  {title} {{Kaon Decays in the Standard Model}},\ }\href
  {https://doi.org/10.1103/RevModPhys.84.399} {\bibfield  {journal} {\bibinfo
  {journal} {Rev. Mod. Phys.}\ }\textbf {\bibinfo {volume} {84}},\ \bibinfo
  {pages} {399} (\bibinfo {year} {2012})},\ \Eprint
  {https://arxiv.org/abs/1107.6001} {arXiv:1107.6001 [hep-ph]} \BibitemShut
  {NoStop}%
\bibitem [{\citenamefont {Egana-Ugrinovic}\ \emph {et~al.}(2020)\citenamefont
  {Egana-Ugrinovic}, \citenamefont {Homiller},\ and\ \citenamefont
  {Meade}}]{Egana-Ugrinovic:2019wzj}%
  \BibitemOpen
  \bibfield  {author} {\bibinfo {author} {\bibfnamefont {D.}~\bibnamefont
  {Egana-Ugrinovic}}, \bibinfo {author} {\bibfnamefont {S.}~\bibnamefont
  {Homiller}},\ and\ \bibinfo {author} {\bibfnamefont {P.}~\bibnamefont
  {Meade}},\ }\bibfield  {title} {\bibinfo {title} {{Light Scalars and the Koto
  Anomaly}},\ }\href {https://doi.org/10.1103/PhysRevLett.124.191801}
  {\bibfield  {journal} {\bibinfo  {journal} {Phys. Rev. Lett.}\ }\textbf
  {\bibinfo {volume} {124}},\ \bibinfo {pages} {191801} (\bibinfo {year}
  {2020})},\ \Eprint {https://arxiv.org/abs/1911.10203} {arXiv:1911.10203
  [hep-ph]} \BibitemShut {NoStop}%
\bibitem [{\citenamefont {Shinohara}(2020)}]{Shinohara:2020brf}%
  \BibitemOpen
  \bibfield  {author} {\bibinfo {author} {\bibfnamefont {S.}~\bibnamefont
  {Shinohara}} (\bibinfo {collaboration} {KOTO}),\ }\bibfield  {title}
  {\bibinfo {title} {{Search for the rare decay $K_L \to \pi^0 \nu \bar \nu$ at
  J-PARC KOTO experiment}},\ }\href
  {https://doi.org/10.1088/1742-6596/1526/1/012002} {\bibfield  {journal}
  {\bibinfo  {journal} {J. Phys. Conf. Ser.}\ }\textbf {\bibinfo {volume}
  {1526}},\ \bibinfo {pages} {012002} (\bibinfo {year} {2020})}\BibitemShut
  {NoStop}%
\bibitem [{\citenamefont {Kitahara}\ \emph {et~al.}(2020)\citenamefont
  {Kitahara}, \citenamefont {Okui}, \citenamefont {Perez}, \citenamefont
  {Soreq},\ and\ \citenamefont {Tobioka}}]{Kitahara:2019lws}%
  \BibitemOpen
  \bibfield  {author} {\bibinfo {author} {\bibfnamefont {T.}~\bibnamefont
  {Kitahara}}, \bibinfo {author} {\bibfnamefont {T.}~\bibnamefont {Okui}},
  \bibinfo {author} {\bibfnamefont {G.}~\bibnamefont {Perez}}, \bibinfo
  {author} {\bibfnamefont {Y.}~\bibnamefont {Soreq}},\ and\ \bibinfo {author}
  {\bibfnamefont {K.}~\bibnamefont {Tobioka}},\ }\bibfield  {title} {\bibinfo
  {title} {{New physics implications of recent search for $K_L \to \pi^0
  \nu\bar{\nu}$ at KOTO}},\ }\href
  {https://doi.org/10.1103/PhysRevLett.124.071801} {\bibfield  {journal}
  {\bibinfo  {journal} {Phys. Rev. Lett.}\ }\textbf {\bibinfo {volume} {124}},\
  \bibinfo {pages} {071801} (\bibinfo {year} {2020})},\ \Eprint
  {https://arxiv.org/abs/1909.11111} {arXiv:1909.11111 [hep-ph]} \BibitemShut
  {NoStop}%
\bibitem [{\citenamefont {Ahn}\ \emph {et~al.}()\citenamefont {Ahn} \emph
  {et~al.}}]{Ahn:2020opg}%
  \BibitemOpen
  \bibfield  {author} {\bibinfo {author} {\bibfnamefont {J.}~\bibnamefont
  {Ahn}} \emph {et~al.} (\bibinfo {collaboration} {KOTO}),\ }\bibfield  {title}
  {\bibinfo {title} {{Study of the $K_L \!\to\! \pi^0 \nu \overline{\nu}$ decay
  at the J-PARC KOTO experiment}},\ }\href@noop {} {\ }\Eprint
  {https://arxiv.org/abs/2012.07571} {arXiv:2012.07571 [hep-ex]} \BibitemShut
  {NoStop}%
\bibitem [{\citenamefont {Caprini}\ \emph {et~al.}(2020)\citenamefont {Caprini}
  \emph {et~al.}}]{Caprini:2019egz}%
  \BibitemOpen
  \bibfield  {author} {\bibinfo {author} {\bibfnamefont {C.}~\bibnamefont
  {Caprini}} \emph {et~al.},\ }\bibfield  {title} {\bibinfo {title} {{Detecting
  gravitational waves from cosmological phase transitions with LISA: an
  update}},\ }\href {https://doi.org/10.1088/1475-7516/2020/03/024} {\bibfield
  {journal} {\bibinfo  {journal} {JCAP}\ }\textbf {\bibinfo {volume} {03}},\
  \bibinfo {pages} {024}},\ \Eprint {https://arxiv.org/abs/1910.13125}
  {arXiv:1910.13125 [astro-ph.CO]} \BibitemShut {NoStop}%
\end{thebibliography}%


\end{document}